# Probing Out-of-Plane Charge Transport in Black Phosphorus with Graphene-Contacted Vertical Field-Effect Transistors

*Junmo Kang[1,†], Deep Jariwala[1,†], Christopher R. Ryder[1], Spencer A. Wells[1], Yongsuk Choi[1,2,3], Euyheon Hwang[2,4], Jeong Ho Cho[1,2,3], Tobin J. Marks[1,5], and Mark C. Hersam[1,5,6],\**

[1]Department of Materials Science and Engineering, Northwestern University, Evanston, Illinois 60208, USA

[2]SKKU Advanced Institute of Nanotechnology (SAINT), Sungkyunkwan University, Suwon 440-746, Republic of Korea

[3]School of Chemical Engineering, Sungkyunkwan University, Suwon 440-746, Republic of Korea

[4]Department of Physics, Sungkyunkwan University, Suwon 440-746, Republic of Korea

[5]Department of Chemistry, Northwestern University, Evanston, Illinois 60208, USA

[6]Department of Electrical Engineering and Computer Science, Northwestern University, Evanston, Illinois 60208, USA

†These authors contributed equally.

*E-mail: m-hersam@northwestern.edu

**Abstract**

Black phosphorus (BP) has recently emerged as a promising narrow band gap layered semiconductor with optoelectronic properties that bridge the gap between semi-metallic graphene and wide band gap transition metal dichalcogenides such as $MoS_2$. To date, BP field-effect transistors have utilized a lateral geometry with in-plane transport dominating device characteristics. In contrast, we present here a vertical field-effect transistor geometry based on a graphene/BP van der Waals heterostructure. The resulting device characteristics include high on-state current densities (> 1600 $A/cm^2$) and current on/off ratios exceeding 800 at low temperature. Two distinct charge transport mechanisms are identified, which are dominant for different regimes of temperature and gate voltage. In particular, the Schottky barrier between graphene and BP determines charge transport at high temperatures and positive gate voltages, whereas tunneling dominates at low temperatures and negative gate voltages. These results elucidate out-of-plane electronic transport in BP, and thus have implications for the design and operation of BP-based van der Waals heterostructures.

The recent successful exfoliation of semiconducting black phosphorus (BP) has fueled intense investigation of its properties in device applications.[1,2,3] Of particular interest has been lateral field-effect transistors (FETs) that exploit in-plane charge transport in BP.[4-7] Early reports described in-plane mobility values for exfoliated BP between ~100 and ~1,000 $cm^2V^{-1}s^{-1}$ at room temperature, depending on sample quality and crystallographic orientation.[1,8] Variable-temperature characterization further indicated phonon-limited mobility at room temperature[9] and impurity-limited mobility below 100 K,[10,11] which suggests a diffusive, band-like transport mechanism for in-plane BP FETs.[1,12] However, the BP electronic structure features a lone pair of electrons at each P atom, which not only can delocalize in-plane but can also interact strongly with out-of-plane atoms.[13-15] Consequently, BP has the potential to exhibit higher out-of-plane conductivities[16] compared to transition metal dichalcogenides (TMDCs) where the chalcogen atoms are largely inactive with respect to out-of-plane carrier transport.[17,18]

Here, vertical field-effect transistors (VFETs) are fabricated from graphene/BP van der Waals heterostructures in order to explore out-of-plane charge transport in BP. Charge transport is explored across different regimes of temperature and carrier density, resulting in the identification of two regimes with distinct transport mechanisms. At low carrier densities (positive gate voltages) and high temperatures, thermionic emission over the graphene/BP Schottky barrier determines charge transport characteristics, whereas at low temperatures and high carrier densities (negative gate voltages), tunneling through the Schottky barrier is the dominant mechanism. The high on-state current density ($> 1600$ A/cm$^2$) in graphene/BP VFETs further demonstrates that BP has superlative out-of-plane electrical conductivity for a p-type semiconductor, and is comparable to n-type MoS$_2$.[19] Overall, this study reveals BP as leading candidate for van der Waals heterostructures where out-of-plane transport determines device performance metrics.

VFETs were fabricated by mechanically exfoliating BP flakes on top of pre-patterned graphene electrodes on 300 nm thick SiO$_2$/Si substrates (Supporting Information S1), followed by electron beam lithography to define the top Ni/Au contact. The high work function of Ni (5.01 eV) combined with superior adhesion enables effective lift-off and Ohmic contact to the hole-doped BP. Figures 1a and 1b present a schematic illustration and an optical micrograph of a representative BP-VFET. The position of the underlying chemical vapor deposition (CVD) derived 10 μm wide graphene stripe is indicated by the dotted lines. In these devices, CVD graphene acts as an atomically thin electrode that is partially transparent to the gate electric field and thereby allows channel modulation even in a vertically stacked geometry.[19-22] A drain-source bias ($V_{SD}$) is applied between the graphene and top metal electrode (grounded), while a gate-source bias ($V_G$) is applied between the Si substrate and graphene. The channel area for the VFET is defined by the overlapping area between the bottom graphene and top metal electrode, while the channel length is determined by the BP flake thickness as measured by atomic force microscopy (AFM) in Figure 1c. Raman spectroscopy of this graphene/BP vertical heterostructure not only shows the $A_g^1$, $B_{2g}$, and $A_g^2$ peaks of BP at 364, 440, and 469 cm$^{-1}$, but also the G and 2D peaks of graphene at 1594 and 2683 cm$^{-1}$ (Figure 1d).[8,23] The Raman peak shapes and positions are consistent with pristine BP and graphene, which confirms no significant perturbation to either material following the formation of the van der Waals heterostructure.

A well-known chemical property of BP is its reactivity in the presence of water, oxygen, and light,[24] implying that BP-based electronic devices often suffer irreversible degradation in ambient conditions.[25, 26] Therefore, the present BP-VFET devices were measured under high vacuum ($<5 \times 10^{-5}$ Torr) immediately following fabrication. Furthermore, based on our earlier studies,[25] best practices were adopted to minimize ambient exposure during fabrication. X-ray photoelectron spectroscopy (XPS) on exfoliated BP flakes following this protocol showed no evidence of phosphorus oxide (see Supporting Information S2). Output characteristics ($J_{SD}$-$V_{SD}$ curves) of a representative device are shown in Figure 2a. The measured current is normalized by the channel area defined by the top metal electrode. The output characteristics exhibit current modulation as a function of gate voltage. In particular, the current density increases with negative $V_G$ values, which indicates *p*-type semiconductor response as expected for BP.[1,8,23,25-27] While the output characteristics are highly linear and symmetric at negative $V_G$ values, they progressively become non-linear and asymmetric for positive $V_G$ values, suggesting a transition from an Ohmic contact to a Schottky contact as will be discussed in more detail later. The transfer characteristics ($J_{SD}$-$V_G$ curves) of the BP-VFETs further confirm the *p*-type polarity of the device as shown in Figure 2b. A high on current density of ~1,600 A/cm$^2$ is observed in the BP-VFETs (Supporting Information S3), which is comparable to previously reported values in *p*-type organic and *n*-type TMDC vertical transistors at similar

$V_D$ values.[16,19-21,28]

At room temperature, the BP-VFETs exhibit an on/off current ratio of ~42 with both the on and off current densities showing a strong bias voltage dependence that suggests barrier-limited transport at room temperature as schematically illustrated in Figure 2c.[19,21,22,28-32] The vertical transistor structure is gate/SiO$_2$/graphene/BP/metal, where the transport is dominated by the Schottky barrier formed at the graphene/BP interface. As $V_G$ transitions from negative to positive values, the Fermi level rises from the top of the BP valence band below the Dirac point in graphene to the center of the BP band gap, resulting in a finite Schottky barrier, to the bottom of the BP conduction band. As $V_G$ becomes increasingly more positive, the barrier height is again reduced to negligible values, corresponding to a rise in current density, once $V_G$ > 40 V, as seen in Figure 2b. The minimum current density occurs at $V_G$ values where the Schottky barrier height is maximized and thermionic emission over the barrier is most limited, hindering charge transport. Despite the barrier, it is worth noting that the off current density in BP-VFETs is relatively high compared to other reported VFETs[19,28] as well as conventional/lateral BP FETs[1,25] (see Supporting Information S4 for direct comparison of lateral and vertical devices on the same BP flake). This relatively high off current density at room temperature can be attributed to the smaller band gap and higher doping level in BP[16,33] compared to TMDCs, which implies shorter depletion widths and lower barrier heights. In addition, the presence of defect and trap states in the gap of BP may provide additional charge transport pathways that are insensitive to gating for ultrashort channel lengths as in the case of VFETs. The lack of saturation in output characteristics is another notable feature in VFETs. Given the ambipolar nature of BP, current saturation under channel pinch-off is limited in a manner that is analogous to ambipolar carbon nanotubes[34] and graphene.[35] In addition, the effects of velocity saturation are not observed in these devices due to their limited range of $V_{SD}$ (attempts to go to higher $V_{SD}$ values resulted in irreversible device failure).

To further elucidate the charge transport mechanisms in BP-VFETs, temperature dependent transfer characteristics of representative devices were acquired from 300 K down to 30 K in steps of 30 K. Figure 3a shows the temperature dependent transfer characteristics ($I_{SD}$-$V_G$ curves) of a BP-VFET (see Supporting Information S5 for output characteristics). The on-state current ($V_G$ = -60 V) shows relatively weak temperature dependence as opposed to the off-state current at $V_G$ ~ 50 V, which possesses an exponential reduction with falling temperatures, resulting in an increased on/off current ratio that saturates at ~800 (Figure 3b). This observation suggests that a thermally activated mechanism dominates charge transport in the off-state at low temperatures. Overall, carrier transport in BP-VFETs can be divided into three distinct regimes depending on gate voltage and temperature with different mechanisms dominating in each regime. To quantify these mechanisms, two models are employed: a tunneling model with weak temperature dependence and a thermionic emission model with exponential temperature dependence as described by the equations below:[29,36]

$$I(V, T) = I(V, 0)[1+(\pi c(V)k_B T)^2/6] \quad (1)$$

$$I_{sat} = AA^*T^2 \exp(qV_{bi}/k_B T) \quad (2)$$

where $c(V)$ is a constant related to the tunneling barrier of the junction, $I_{sat}$ is the saturation current extrapolated from the logarithmic output characteristics[19] (see Supporting Information S6), $A$ is the area of junction, $A^* = 4\pi q m^* k_B^2 h^{-3}$ is the effective Richardson constant, $m^*$ is the carrier effective mass, $h$ is the Plank constant, $k_B$ is the Boltzmann constant, $V_{bi}$ is the Schottky barrier height, $q$ is the fundamental unit of charge, and $T$ is temperature.

At low $V_G$ (< 10 V), the current exhibits a weak dependence across the full range of temperatures. Tunneling is thus likely to be the dominant transport mechanism in this regime. The BP thickness in the VFETs is sufficiently large (> 10 nm) so as to have negligible direct (source to drain) tunneling. However, tunneling can occur in multiple steps *via* impurity or defect states.[37,38] (see Supporting Information S7 for an estimate of the defect density). A fit of the temperature dependent current data to eqn. (1) (Figure 3c) confirms the dominant tunneling mechanism in this regime. The same data can be fit reasonably well to a hopping transport model (see Supporting Information S8), which also indicates the presence of impurities and defects. At high $V_G$ (> 10 V) and high temperatures, the current varies exponentially with temperature, agreeing well with the thermionic emission model (Figure 3d, dashed lines), whereas at high $V_G$ and low temperatures (< 180 K), the current once again closely resembles what is expected for the tunneling model (Figure 3e, solid lines). This transition in charge transport mechanism as a function of temperature is further illustrated in an Arrhenius plot of $\ln(I_{sat}/T^2)$ versus $q/k_B T$ (Figure 3d). All plots for $V_G$ > 20 V show two distinct slopes corresponding to distinct mechanisms for charge transport. Specifically, these curves transition from a linear dependence (large negative slope) to a nearly temperature independent behavior (negligible slope) at lower temperature (below 180 K), further suggesting a transition from thermionic emission to tunneling. The Schottky barrier heights extracted from the Arrhenius plots are shown in Figure 3f as a function of gate voltage. The barrier height peaks at $V_G \sim 50$ V, which coincides with the $V_G$ of lowest observed off current values. However, the off currents in these BP-VFETs (few nA) are still much higher than in TMDC VFETs (few pA),[19,28,39] which can be explained by direct quantum mechanical tunneling through the depletion region of the Schottky junction. Tunneling through the Schottky barrier has been reported for TMDC FETs[40] since the barrier width is limited by the semiconductor thickness. In the case of graphene/BP, the barrier height and width are expected to be lower and narrower, respectively, due to the smaller gap and higher doping level in BP compared to TMDCs, thereby leading to larger tunneling currents.

The above discussions elucidate the origin of high off-currents and consequently low on/off ratios in BP-VFETs. However, VFETs based on two-dimensional materials are intrinsically short channel devices and thus short channel effects should also be considered. To investigate these effects, we first fabricated VFETs with varying BP thickness. The on-state current density in BP-VFETs remains insensitive to the BP thickness (Figure 4a) suggesting that the BP channel resistance is negligible compared to the graphene/BP contact resistance (see Supporting Information S9). In contrast, the on/off ratio decreases with increasing BP thickness (Figure 4b), which is consistent with increased screening of the gate electric field with increasing thickness of the BP flake. A prominent short channel effect is the gradual loss of gate control over the channel.[41,42] To improve gate control, the capacitive coupling from the gate can be increased by reducing dielectric thickness and using high-k materials.[43,44] To determine if the diminished gate coupling is limiting the on/off ratios in BP-VFETs additional devices were fabricated on 50 nm thick $AlO_x$ dielectric grown by atomic layer deposition (Figure 4c and 4d, see Supporting Information S10 for more details). Despite the significant increase in capacitive gate coupling in this case ($1.24 \times 10^{-7}$ F cm$^{-2}$, for the 50 nm thick $AlO_x$ layer),[45] no significant improvement in gate control is observed since the on/off ratios and on current densities remain in the same range while the operating voltage window is reduced from 120 V (for 300 nm $SiO_2$) to 16 V (for 50 nm $AlO_x$). In earlier studies of TMDC VFETs, a smaller than expected on/off ratio was also attributed to Fermi level pinning at the graphene/semiconductor interface due to trap states in the semiconductor gap region.[28]

However, in the case of BP-VFETs, the Fermi level moves through the entire gap region since a small onset of *n*-type transport is observed in the transfer curve (both in the case of 300 nm $SiO_2$ and 50 nm $AlO_x$), confirming little to no pinning of the Fermi level at the graphene/BP interface. Based on the above discussion, it is apparent that short channel effects are not likely to be the root cause of the high off-currents and consequently low on/off ratios in BP VFETs, particularly in the range of BP thickness and gate capacitance investigated here.

In conclusion, BP-VFETs have been fabricated using graphene as a bottom electrode in order to study out-of-plane charge transport in BP. High off-currents are found to limit the on/off ratios in the resulting BP-VFETs. Furthermore, thermionic emission over the graphene/BP Schottky barrier is identified as the source of the high off currents at high temperatures, while tunneling through the Schottky barrier results in relatively high off currents at low temperatures. The graphene/BP interface is also observed to have high electronic quality, which allows unimpeded displacement of the Fermi level through the entire gap. Overall, the above results suggest that BP has superlative out-of-plane conductivity but suffers from poor gate modulation in the VFET geometry due to its small band gap and presence of impurity/defect states. These results thus motivate ongoing efforts to improve BP material quality as well as the integration of BP into vertical van der Waals heterostructures.

*Methods*

**BP exfoliation and device fabrication.** Mechanical exfoliation of BP crystals was carried out in an Ar glove box to minimize $O_2$ and $H_2O$ degradation. The BP flakes were directly exfoliated onto patterned graphene/$SiO_2$/Si or patterned graphene/$AlO_x$/Si substrates. The top electrodes in the BP-VFETs and the contacts to the underlying graphene stripes were defined using electron beam lithography. The metallization for the contacts was Ni/Au (20/40 nm thick).

**Characterization.** AFM images were obtained in tapping mode using an Asylum Cypher S system. Raman spectroscopy with excitation wavelengths of 514 nm (Renishaw) and 532 nm (Horiba) was used to characterize the quality of the graphene and BP. All electrical measurements were performed under high vacuum ($< 5 \times 10^{-5}$) in a Lake Shore CRX 4K probe station using source-meter units (Keithley 2400).

**ASSOCIATED CONTENT**

**Supporting Information**:

Additional details on device fabrication and materials characterization as well as supplemental electrical data and its analysis accompany this paper and are available free of charge *via* the Internet at http://pubs.acs.org.

**AUTHOR INFORMATION**

**Corresponding author:**

*E-mail: m-hersam@northwestern.edu

**NOTES:**

**Competing financial interests**: The authors declare no competing financial interests.

**ACKNOWLEDGEMENTS**

This research was supported by the NSF Materials Research Science and Engineering Center (MRSEC) of Northwestern University (DMR-1121262), the NSF EFRI 2-DARE program (NSF EFRI-143510), and the Office of Naval Research (N00014-14-1-0669). J.K. acknowledges support from the Postdoctoral Research Program of Sungkyunkwan University. D.J. acknowledges additional support from an SPIE education scholarship and IEEE DEIS fellowship. This work made use of the Northwestern University Micro/Nano Fabrication Facility (NUFAB) as well as the Northwestern University Atomic and Nanoscale Characterization Experimental Center (NUANCE), which has received support from the MRSEC (NSF DMR-1121262), State of Illinois, and Northwestern University.

**Figures**

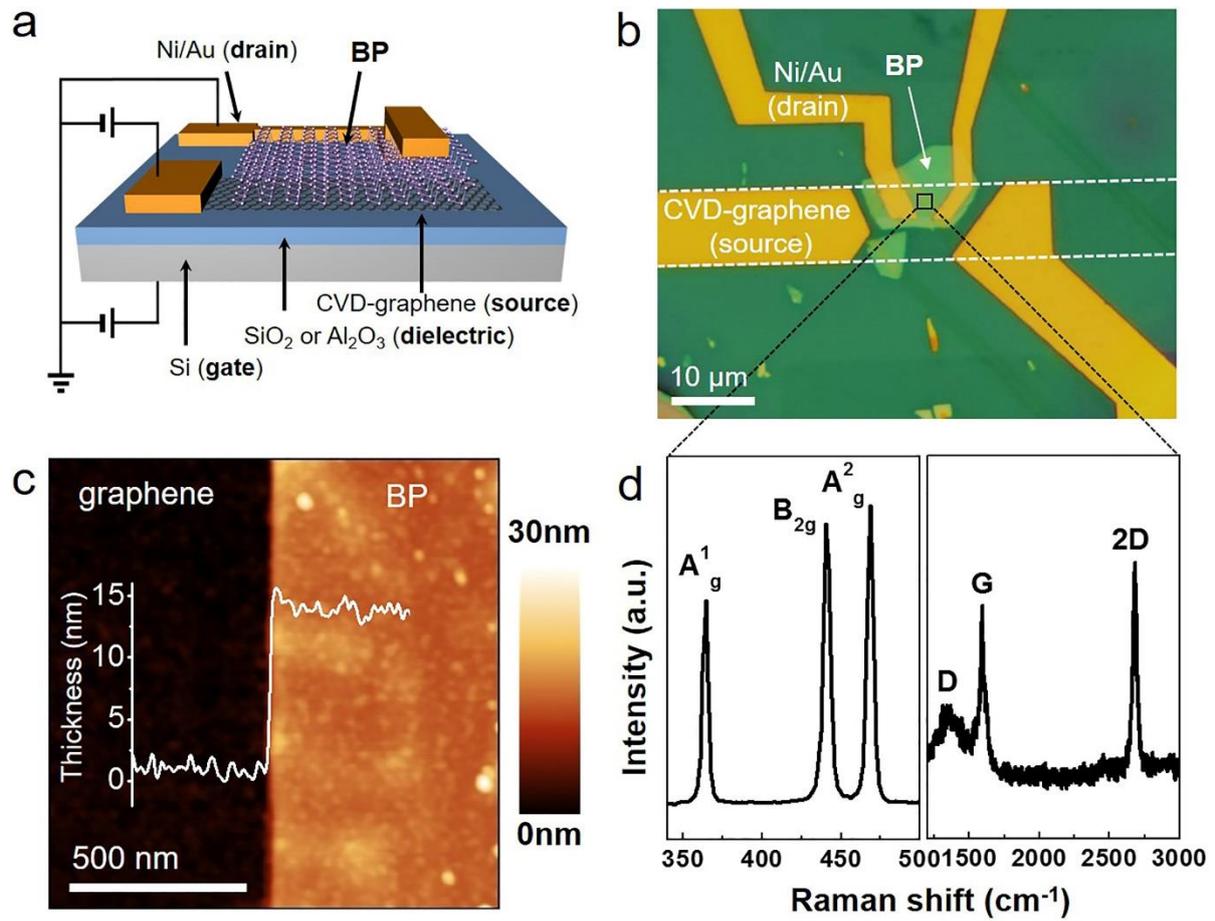

**Figure 1.** Graphene/BP vertical field-effect transistor (VFET). (a) Schematic illustration of the BP-VFET. (b) Optical microscopy image of the fabricated device. The channel area of the BP-VFET is 8.4 μm$^2$ in this case. (c) AFM image at the edge of the BP flake. Inset: cross-sectional height profile of the BP flake on the graphene/SiO$_2$/Si substrate. (d) Raman spectrum of the graphene/BP vertical heterostructure.

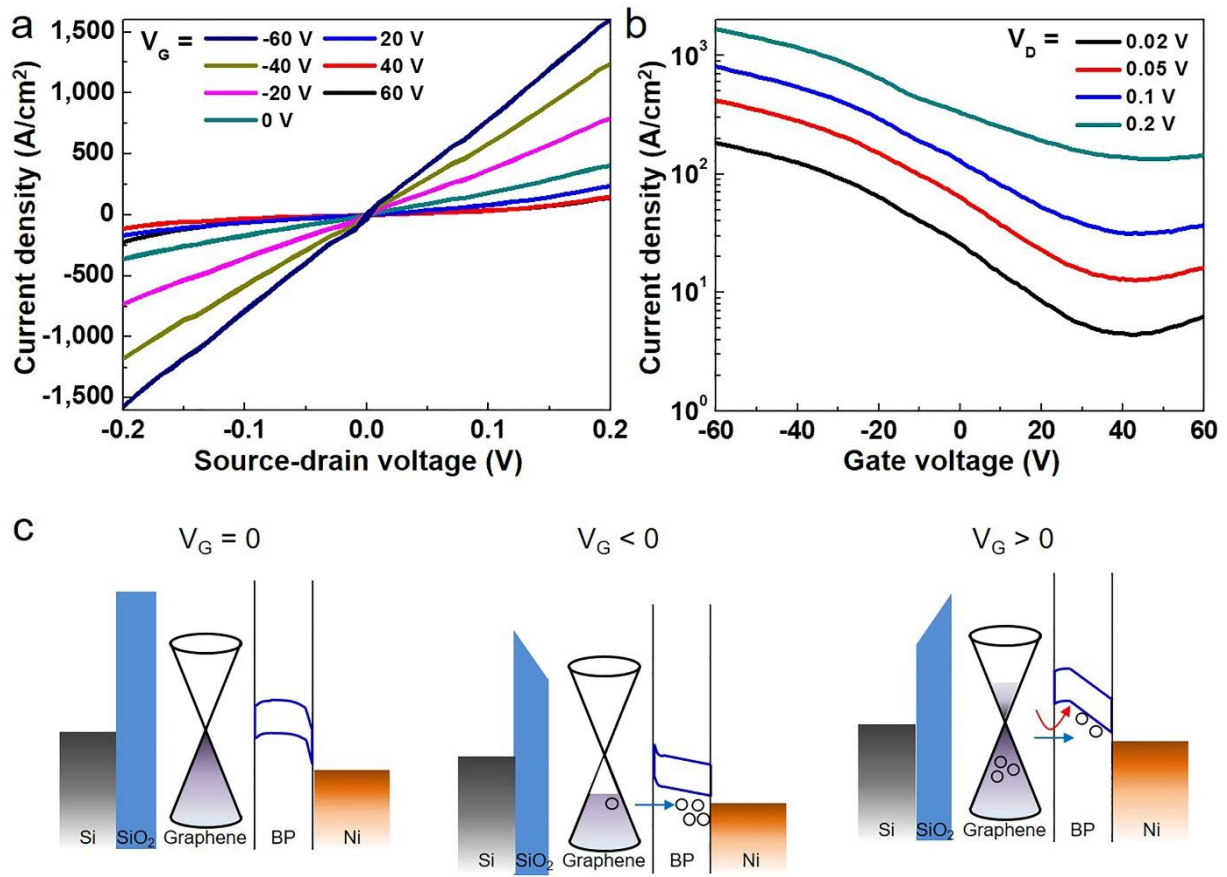

**Figure 2.** Room temperature transport characteristics of the BP-VFET. (a) $J_{SD}$-$V_{SD}$ output characteristics for different values of $V_G$. (b) Semi-logarithmic $J_{SD}$-$V_G$ transfer characteristics for different values of $V_{SD}$. (c) Schematic band structures at zero gate voltage, negative gate voltage, and positive gate voltage.

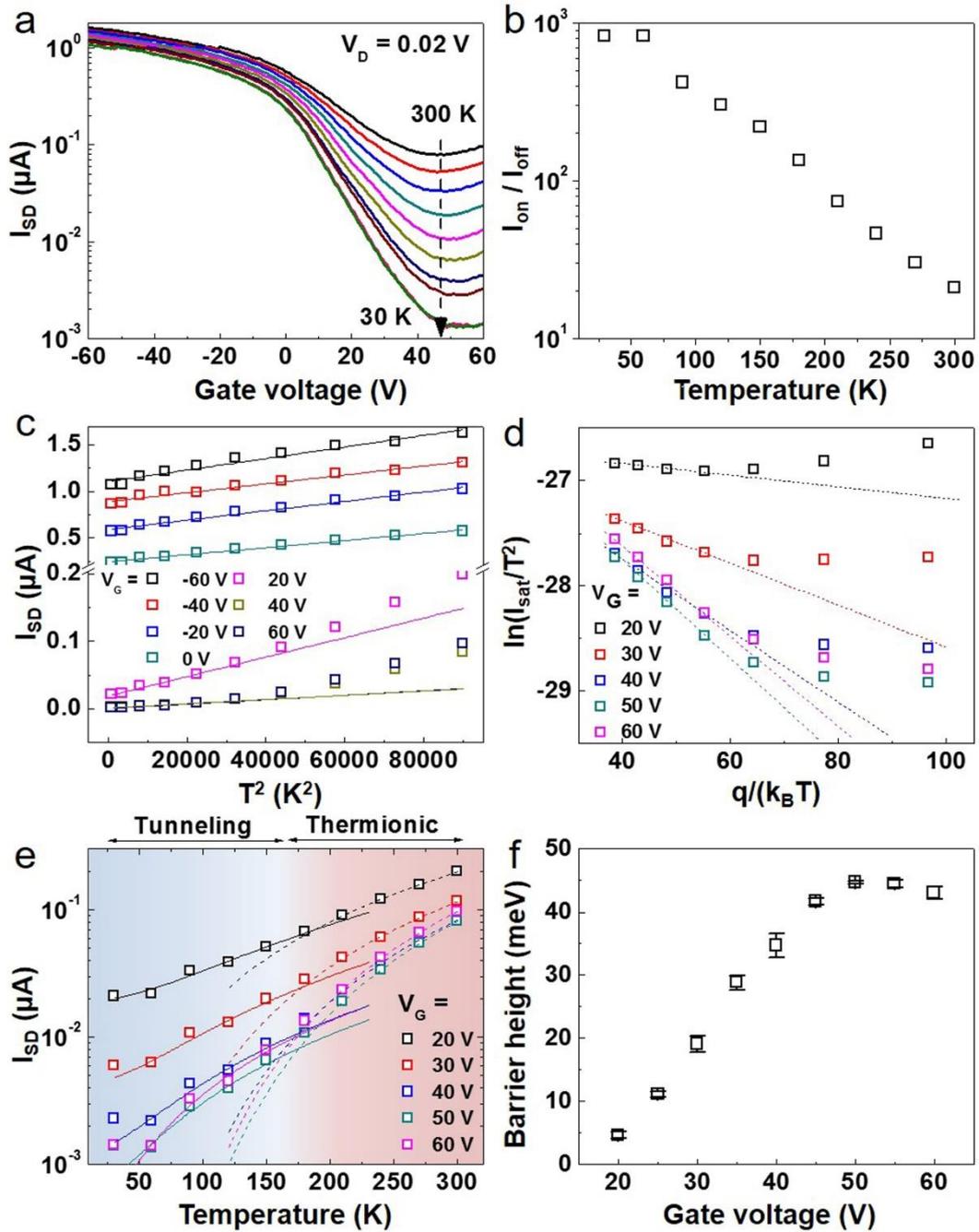

**Figure 3.** Temperature dependent charge transport in the BP-VFET. (a) Semi-logarithmic $I_{SD}$-$V_G$ transfer characteristics at different temperatures ranging from 300 K to 30 K in 30 K steps. (b) On/off ratio as a function of temperature. (c) Plot of $I_{SD}$ as a function of $T^2$ showing the best fits to the experiment data over the temperature range of 180 K to 30 K. (d) Temperature dependent saturation current at different gate voltages from 20 V to 60 V in 10 V steps, showing the best fits to the experiment data over the temperature range of 300 K to 210 K. (e) Temperature dependence of $I_{SD}$ at different gate voltages from 20 V to 60 V in 10 V steps. (f) Corresponding Schottky barrier heights obtained from the slope of (d).

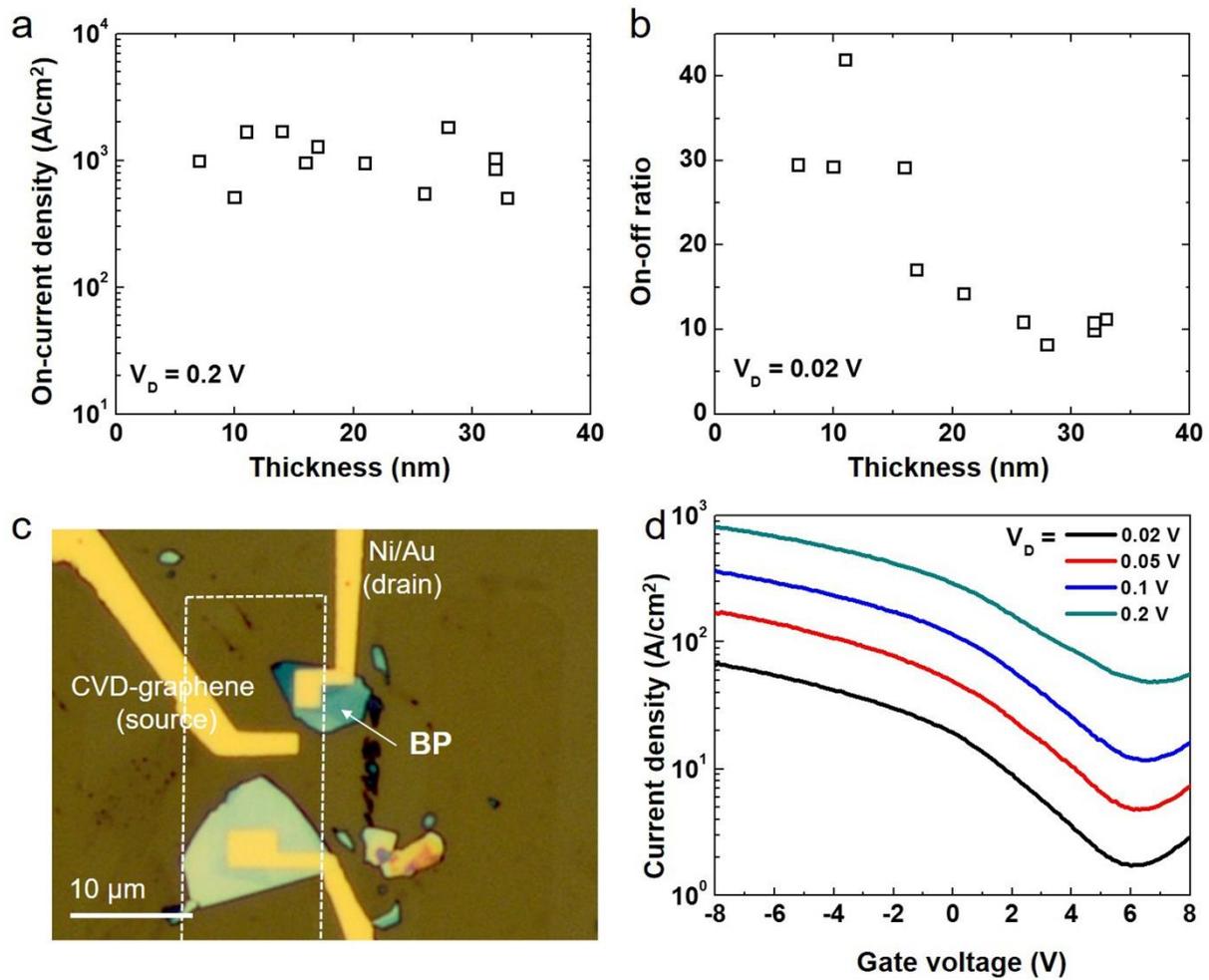

**Figure 4.** (a) BP-VFET on-current density as a function of BP thickness at a source-drain bias of 0.2 V. (b) BP-VFET on-off ratio as a function of BP thickness at a source-drain bias of 0.02 V. (c) Optical micrograph of a BP-VFET fabricated on 50 nm thick $AlO_x$. The BP-VFET channel area is 6 $\mu m^2$ in this case. (d) Semi-logarithmic transfer characteristics of the BP-VFET from (c) for different values of source-drain bias.

# Supporting Information

## Probing Out-of-Plane Charge Transport Black Phosphorus with Graphene-Contacted Vertical Field-Effect Transistors

*Junmo Kang[1,†], Deep Jariwala[1,†], Christopher R. Ryder[1], Spencer A. Wells[1], Yongsuk Choi[1,2,3], Euyheon Hwang[2,4], Jeong Ho Cho[1,2,3], Tobin J. Marks[1,5], and Mark C. Hersam[1,5,6,]\**

[1]Department of Materials Science and Engineering, Northwestern University, Evanston, Illinois 60208, USA

[2]SKKU Advanced Institute of Nanotechnology (SAINT), Sungkyunkwan University, Suwon 440-746, Republic of Korea

[3]School of Chemical Engineering, Sungkyunkwan University, Suwon 440-746, Republic of Korea

[4]Department of Physics, Sungkyunkwan University, Suwon 440-746, Republic of Korea

[5]Department of Chemistry, Northwestern University, Evanston, Illinois 60208, USA

[6]Department of Electrical Engineering and Computer Science, Northwestern University, Evanston, Illinois 60208, USA

†These authors contributed equally.

*E-mail: m-hersam@northwestern.edu

**S1. Synthesis, patterning, and characterization of graphene**

Monolayer graphene was grown on 25 μm thick Cu foil using chemical vapor deposition (CVD). In particular, a Cu foil was inserted into a 2 inch quartz tube and annealed at 970 ºC for 1 h under 2 sccm $H_2$ flow. Then, a 20 sccm $CH_4$ flow was injected to grow the graphene for 0.5 h. Subsequently, the tube was rapidly cooled down to room temperature under He flow. Poly(methyl methacrylate) (PMMA) was coated on the graphene film on the front side of the Cu foil, after which graphene on the back side was etched away with $O_2$ plasma. The Cu foil was removed by an aqueous solution of 0.1 M ammonium persulfate solution $((NH_4)_2S_2O_8)$. After rinsing three times with deionized water, the floating PMMA/graphene film was placed onto the $SiO_2$/Si or $AlO_x$/Si substrates. The PMMA was then dissolved in acetone. Graphene electrodes were defined by photolithography and oxygen plasma etching. The size of the patterned graphene electrodes was 10 μm × 100 μm.

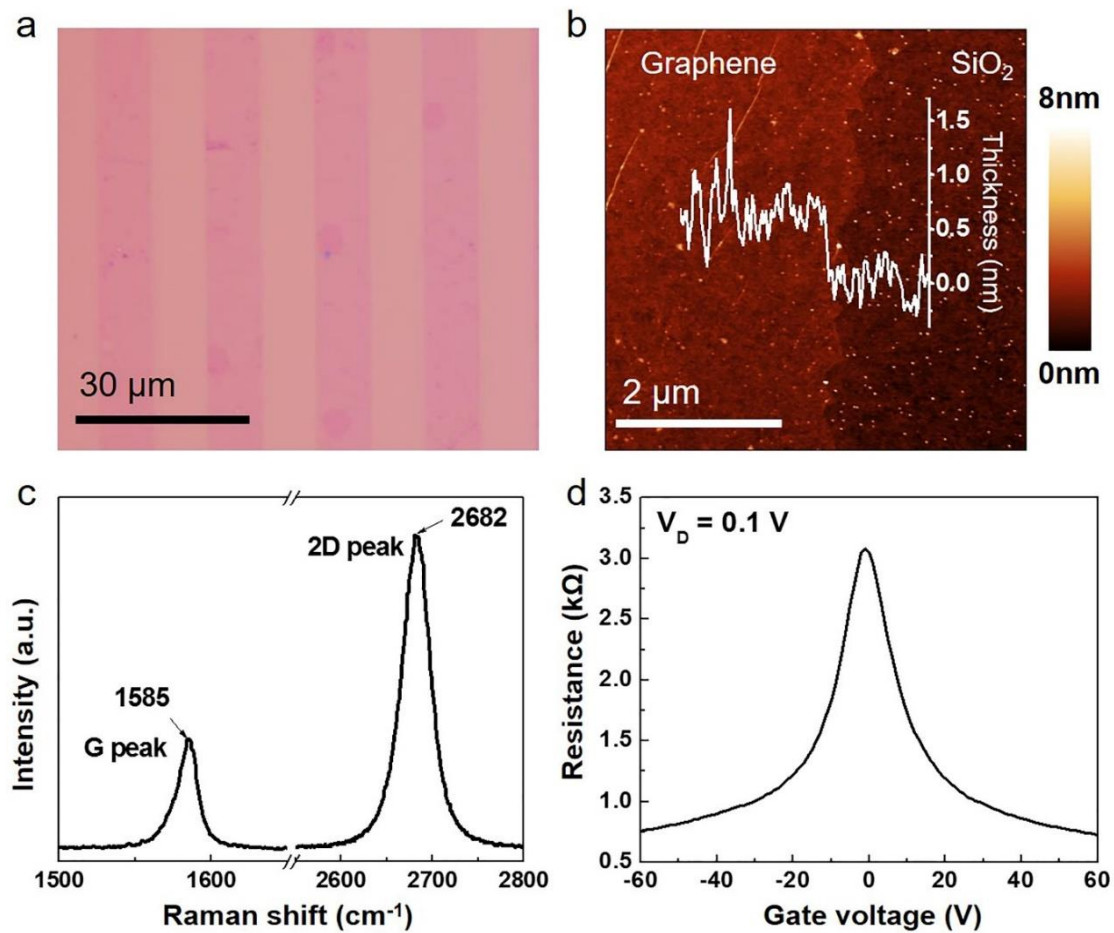

**Figure S1.** Preparation of CVD graphene electrodes on a $SiO_2$/Si substrate. (a) Optical microscopy image of CVD graphene strips with 10 μm width and 100 μm length. (b) Atomic force microscopy image of the edge of a CVD graphene strip. The height profile in the inset shows that the graphene thickness is ~0.4 nm. (c) 514 nm Raman spectrum of CVD graphene on a $SiO_2$/Si substrate. (d) Transfer characteristics of a CVD graphene transistor at a source-drain voltage of 0.1 V. The Dirac point of the CVD graphene electrode shows almost zero gate voltage with a carrier mobility of 1600 $cm^2$/Vs.

**S2. XPS analysis of mechanically exfoliated BP**

The chemical state of pristine BP samples was measured by X-ray photoelectron spectroscopy (XPS). The BP samples were exfoliated onto 300 nm $SiO_2$ substrates in a nitrogen glovebox (< 0.25 ppm oxygen) and immediately transferred to the high vacuum environment of the XPS system. Exposure to ambient conditions was less than 30 seconds. P 2p core level spectra in Figure S2 show a clear BP doublet at ~130 eV and a broad Si satellite peak from the substrate at ~127 eV. Notably absent are any features associated with phosphorus oxide species at 132-134 eV,[1,2] indicating that there is no measurable oxide at the BP surface.

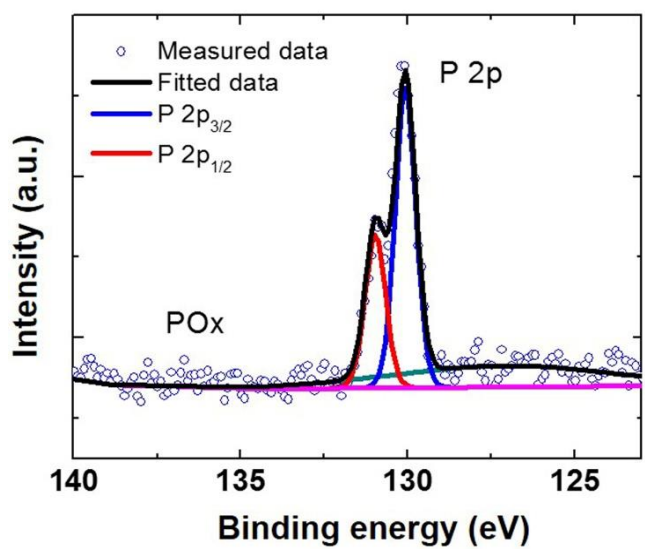

**Figure S2.** P 2p XPS spectrum for a mechanically exfoliated BP flake on a 300 nm thick $SiO_2$ substrate.

## S3. Room temperature characterization of BP-VFETs on a 300 nm SiO$_2$/Si substrate

BP-VFETs were fabricated on a 300 nm SiO$_2$/Si substrate to investigate BP out-of-plane charge transport. The results in Figure S3 supplement Figure 2 of the main manuscript.

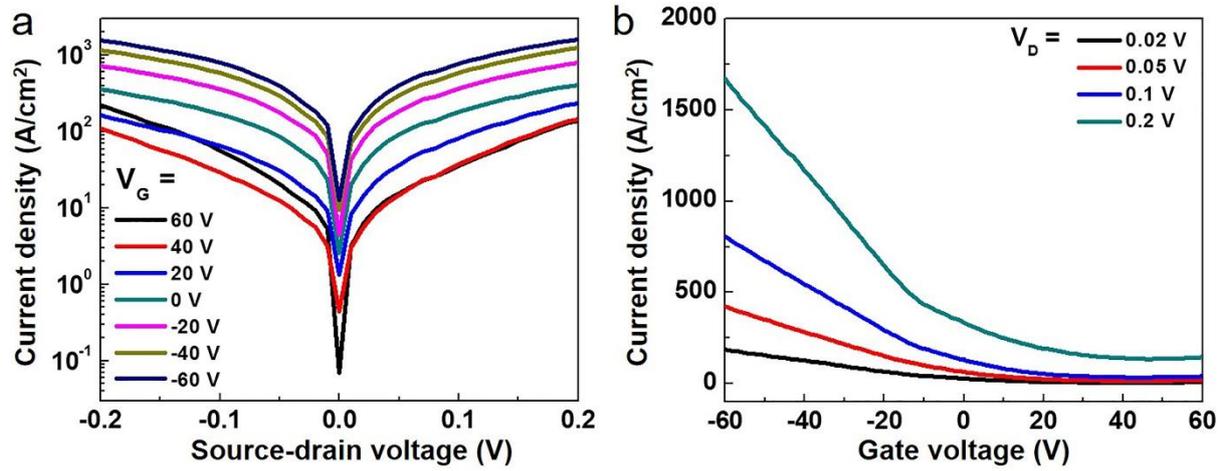

**Figure S3.** Room-temperature characterization of a BP-VFET on a 300 nm SiO$_2$/Si substrate. (a) Semi-logarithmic $J_{SD}$-$V_{SD}$ output characteristics for different values of $V_G$. (b) $J_{SD}$-$V_G$ transfer characteristics for different values of $V_{SD}$.

**S4. Comparison of vertical and lateral BP FETs**

Both vertical FET (VFET) and lateral FET (LFET) devices were fabricated from the same BP flake and measured as shown in Figure S4. Both the VFET and LFET have the same source/drain (metal/graphene) contact. The channel length of the LFET is determined by the distance between the graphene edge and the Ni/Au metal electrode. The transfer characteristics show clear differences with the LFET exhibiting lower off-current and consequently higher on/off ratios.

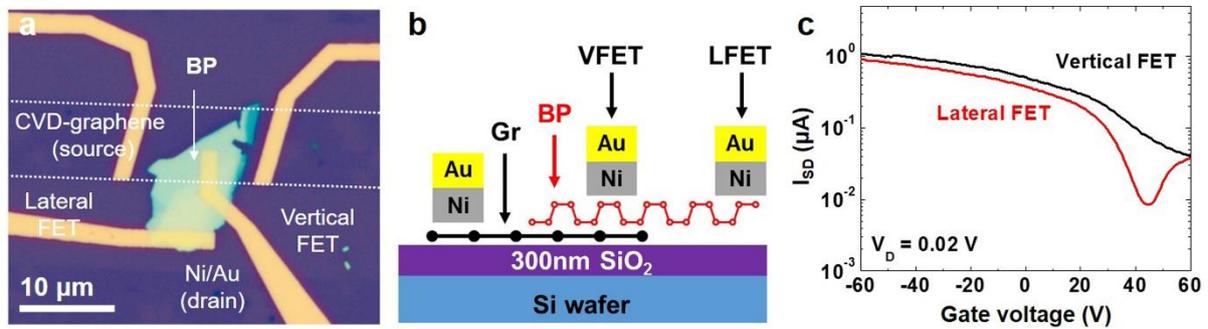

**Figure S4.** VFETs and LFETs based on the same BP flake. (a) Optical microscopy image of the VFET and LFET fabricated from the same BP flake (b) Schematic illustration of the VFET and LFET with bottom graphene electrode and top Ni/Au metal electrode. (c) Semi-logarithmic $I_{SD}$-$V_G$ transfer characteristics of the VFET and LFET at $V_{SD} = 0.02$ V.

**S5. Variable temperature characterization of BP-VFETs on a 300 nm SiO$_2$/Si substrate**

Temperature dependent output characteristics ($J_{SD}$-$V_{SD}$ curves) of a BP-VFET on a 300 nm SiO$_2$/Si substrate at $V_G$ = -60 V and 60 V (on and off state respectively) are shown in Figure S5a and Figure S5b, respectively. While the output characteristics remain linear in the on state (Figure S5a) throughout the temperature range, non-linearities and rectification are observed in the off state (Figure S5b) due to diminishing thermionic emission over the Schottky barrier.

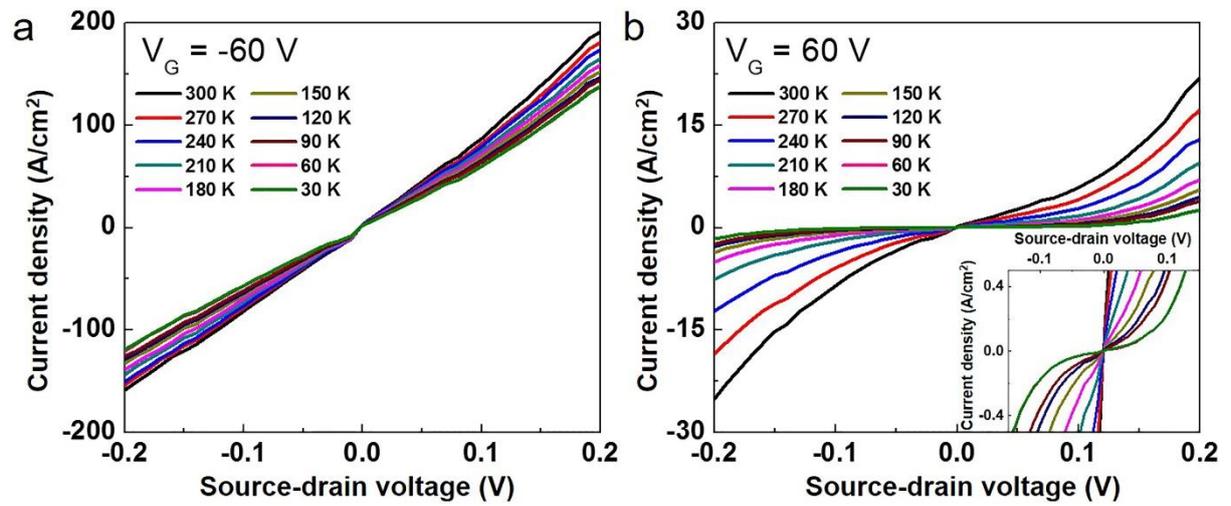

**Figure S5.** $J_{SD}$-$V_{SD}$ output characteristics of a BP-VFET on a 300 nm SiO$_2$/Si substrate for temperatures between 300 K and 30 K at (a) $V_G$ = -60 V and (b) $V_G$ = 60 V. The inset of (b) shows an enlarged plot of $J_{SD}$-$V_{SD}$.

## S6. Arrhenius analysis of the Schottky barrier height

Fits to the Arrhenius equation were performed to extract the Schottky barrier height from the temperature dependent charge transport data. Figure S6a and Figure S6b show the results for $V_G = -60$ V and $V_G = 60$ V, respectively. From the Arrhenius plots of $\ln(I_{sat}/T^2)$ *versus* $q/k_B T$ (Figure S6c), the Schottky barrier height at the graphene/BP interface is extracted. The manuscript discusses this Schottky barrier height in the high gate voltage range ($V_G > 10$ V). However, in the low voltage range ($V_G < 10$), the Schottky barrier height values are negative, ranging from -36 meV to -15 meV (Figure S6d), suggesting an Ohmic contact between the graphene and BP.

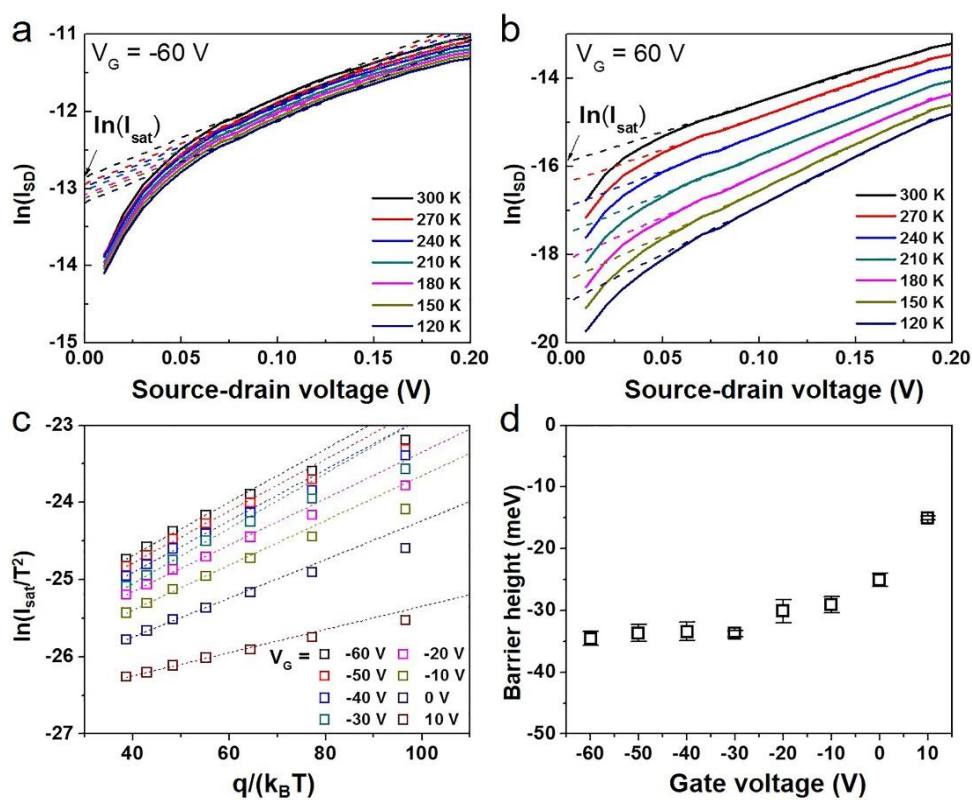

**Figure S6.** Arrhenius analysis of BP-VFETs. Output characteristics of a BP-VFET for temperatures between 300 K and 120 K at (a) $V_G = -60$ V and (b) $V_G = 60$ V. (c) Arrhenius plots at gate voltages between -60 V and 10 V. (d) Corresponding Schottky barrier heights obtained from the slope of (c).

## S7. Interface trap analysis

We have estimated the interface trap density of the lateral BP FET using the following equation:

$$S = \frac{dV_G}{d(\log I_D)} = 2.3\frac{kT}{q}\left[1+\frac{C_d+C_{it}}{C_i}\right] \approx 2.3\frac{kT}{q}\left[1+\frac{C_{it}}{C_i}\right] \quad (1)$$

where $S$ is the sub-threshold swing, $C_d$ is the depletion capacitance, $C_i$ is the dielectric capacitance, and $C_{it}$ is the interface state capacitance (given by $qD_{it}$ where $D_{it}$ is the interface trap charge density and $q$ is electronic charge), $k$ is Boltzmann's constant, and $T$ is temperature in Kelvin. An average value of $D_{it} = 2.11 \times 10^{12}$ cm$^{-2}$/eV was determined using the $S$ values from our lateral BP FETs.

**S8. Variable range hopping model**

The variable range hopping model was considered for both lateral and vertical BP FETs. The lateral FETs show increasing on-state current and transconductance with reducing temperature (Figure S8a), indicating a phonon scattering mechanism and band-like transport. This mobility increase saturates below 100 K (Figure S8a inset) and is limited by charged impurity scattering as observed previously for the cases of graphene[3] and $MoS_2$.[4] The conductance data for the lateral FET fit well with the two-dimensional variable range hopping model (Figure S8b) similar to the case of $MoS_2$, which suggests the presence of traps and impurity states[5,6] in the gap region. In contrast, the vertical FET shows minimal dependence of the on-state current or transconductance on temperature (Figure S8c). Nevertheless, the out-of-plane conductance data also fit to the variable range hopping model, suggesting that traps and impurities play a role in charge transport (Figure S8d).

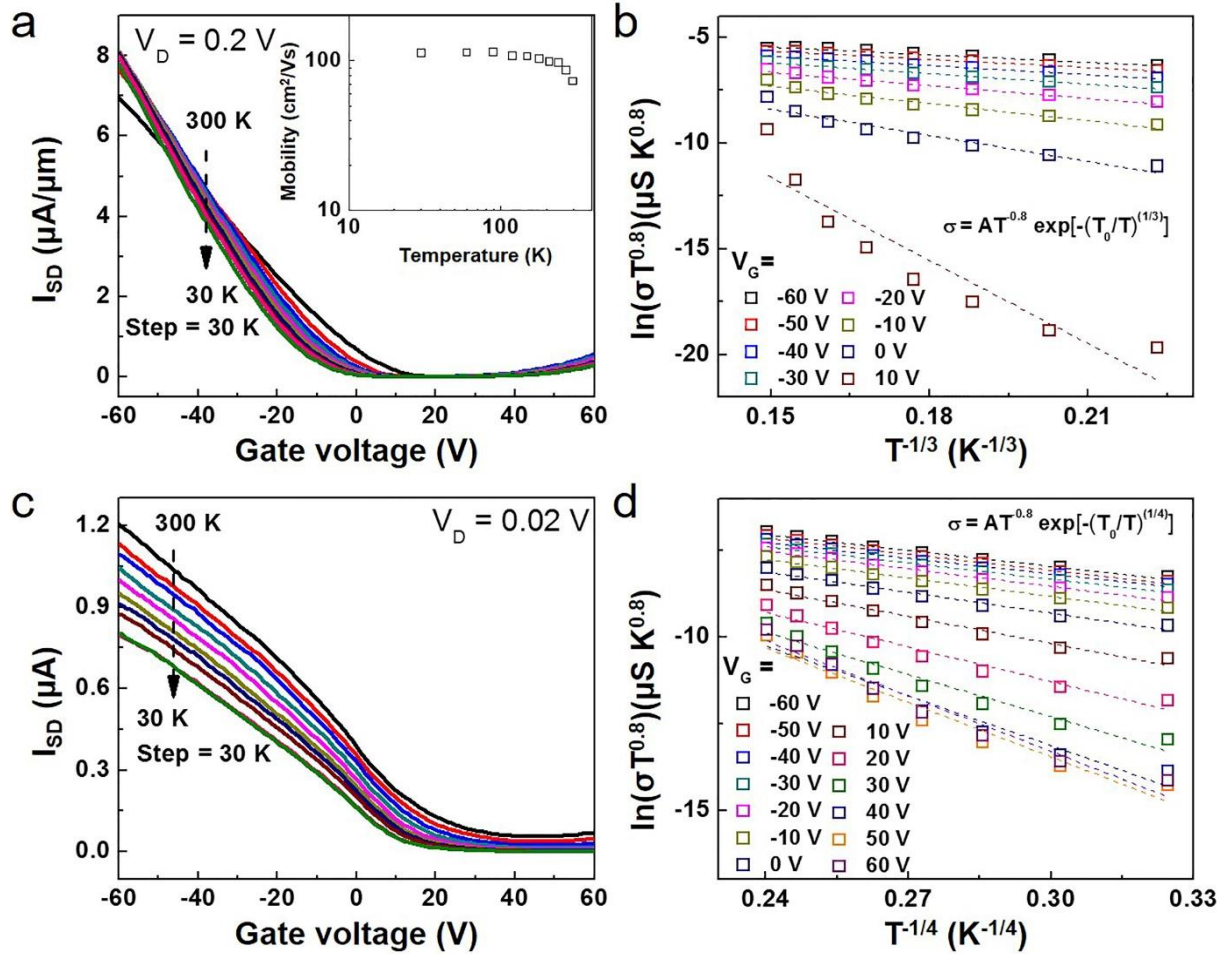

**Figure S8.** Temperature dependent transfer characteristics and conductivity for lateral and vertical BP FETs. (a,c) Source-drain current $I_{SD}$ as a function of gate voltage $V_G$ measured in (a) lateral and (c) vertical BP FETs. The inset of (a) shows temperature dependent field-effect mobility for the lateral BP FET. (b,d) Temperature dependent conductivity and fits using the variable range hopping model for (b) lateral and (d) vertical BP FETs. The dashed lines are the fits to the variable range hopping model.

## S9. Contact resistance characteristics of BP-VFET

The contact resistance per unit area of the graphene/BP Schottky junction was estimated by averaging over three devices with different areas on the same BP flake to avoid the effects of flake-to-flake thickness and defect density variations.[7] Figure S9a shows an optical micrograph of a representative device. Control graphene FETs were also measured to correct for the graphene channel and graphene/metal contact resistance. The average resistance of the BP-VFET is $\sim 10^4$ $\Omega$, which consists of the metal-graphene contact resistance, graphene channel resistance, graphene-BP contact resistance, BP channel resistance, and BP-metal contact resistance. The BP channel ($\sim$10 nm thick) resistance and BP-metal contact resistance are negligible. The graphene/metal contact and the graphene channel resistance together account for only 5% of the total BP-VFET resistance in the off-state. The corrected BP-VFET resistance (red curve) and the graphene/BP contact resistance per unit area (blue curve) are shown in Figure S9b.

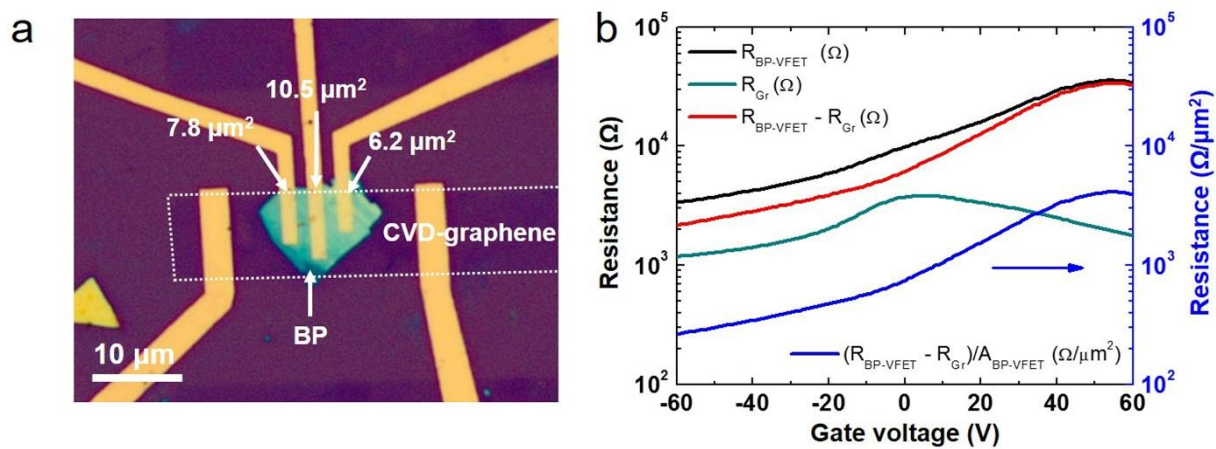

**Figure S9.** (a) Representative optical micrograph of a BP-VFET with varying active areas of the VFET devices. (b) Gate-bias dependence of BP-VFET resistance as-measured (black curve) and corrected (red curve). The blue curve shows the graphene/BP contact resistance, while the green curve shows the resistance of the graphene stripe as a function of gate voltage with a V-shape characteristic of graphene FETs.

## S10. BP-VFETs on 50 nm AlO$_x$/Si substrate

BP-VFETs were also fabricated on 50 nm thick atomic layer deposited AlO$_x$ (1.24 × 10$^{-7}$ F cm$^{-2}$, for the 50 nm thick Al$_2$O$_3$ layer)[8] to investigate the effect of increased gate coupling with reduced dielectric thickness. The gate-dependent output and transfer characteristics in Figure S10 supplement Figure 4c and 4d of the main manuscript.

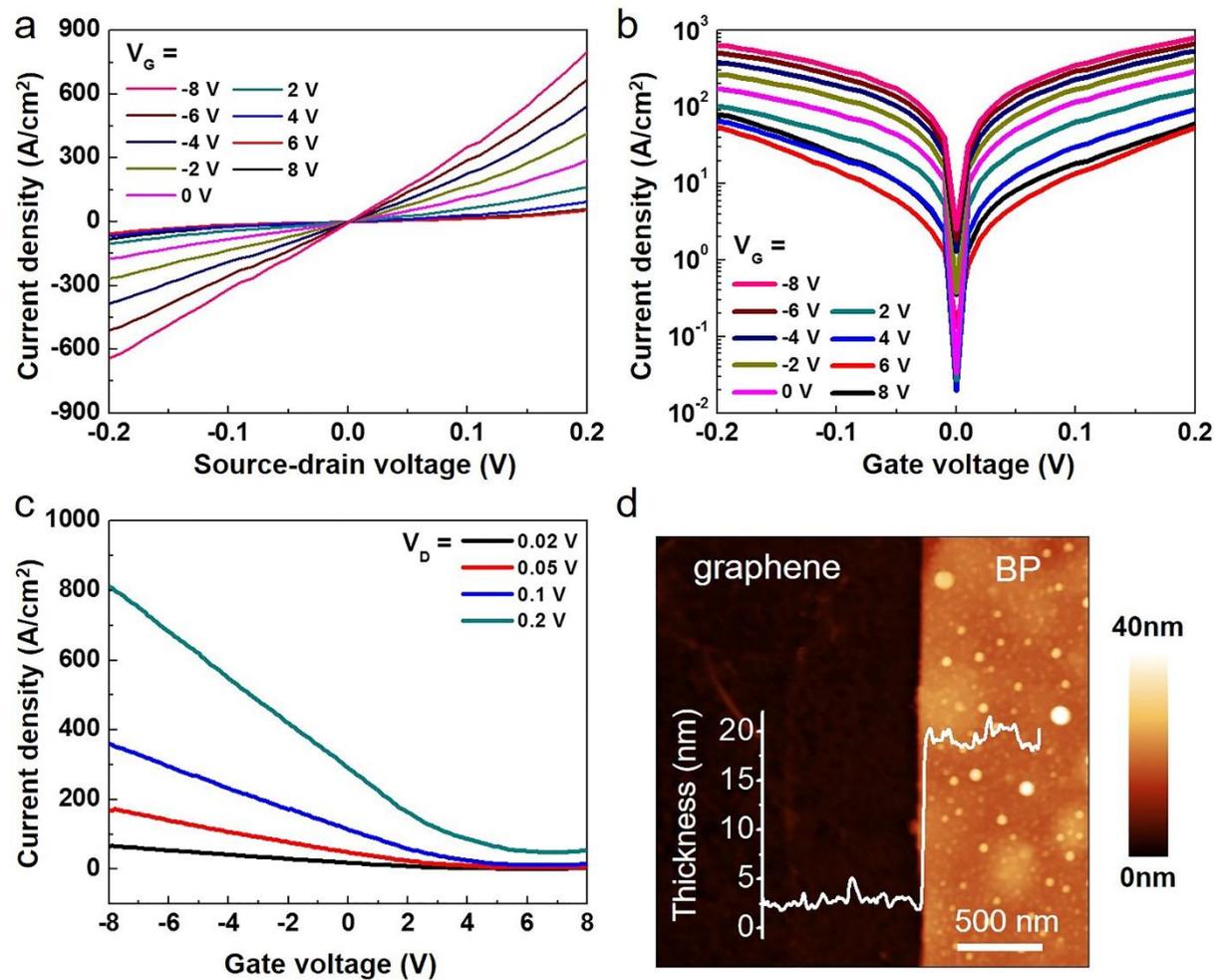

**Figure S10.** Low voltage operation of a BP VFET based on 50 nm thick atomic layer deposited AlO$_x$. (a) Linear scale $J_{SD}$-$V_{SD}$ output characteristics for different values of $V_G$. (b) Semi-logarithmic scale $J_{SD}$-$V_{SD}$ output characteristics for different values of $V_G$. (c) $J_{SD}$-$V_G$ transfer characteristics for different values of $V_{SD}$. (d) AFM image of the edge of the BP flake. The inset shows the height profile of the BP flake on the AlO$_x$ substrate.